\documentclass[a4paper,twocolumn]{IEEEtran}
\usepackage{latexsym}
\usepackage{color}
\usepackage[dvips]{graphicx}
\usepackage{amsmath}
\usepackage{amssymb}
\usepackage[latin1]{inputenc}

\definecolor{Blue}{rgb}{0.2,0.4,0.6}
\definecolor{Red}{rgb}{0.6,0.3,0.3}
\definecolor{Green}{rgb}{0.2,0.6,0.4}
\definecolor{Gray}{gray}{0.5}

\begin{document}
\bibliographystyle{IEEEtran}
\title{Side-Information Coding with Turbo Codes and its Application to Quantum Key Distribution\thanks{We thank Jean Cardinal for many fruitful discussions. This work was supported by the Communauté Française de Belgique under grant ARC 00/05-251 and by the IUAP programme of the Belgian government under grant V-18.}}

\author{Kim-Chi Nguyen\thanks{K.-C. Nguyen, G. Van Assche, and N. J. Cerf are
with the Centre for Quantum Information and Communication, Ecole Polytechnique,
CP 165, Université Libre de Bruxelles, 1050 Brussels, Belgium. (e-mail:
kimchng@ulb.ac.be, gvanassc@ulb.ac.be, ncerf@ulb.ac.be)}, Gilles Van Assche, and Nicolas J. Cerf}

\maketitle

\begin{abstract}
Turbo coding is a powerful class of forward error correcting codes, which can achieve performances close to the Shannon limit. The turbo principle can be applied to the problem of side-information source coding, and we investigate here its application to the reconciliation problem occuring in a continuous-variable quantum key distribution protocol.
\end{abstract}


% ----------------------------------------------------------
%
\section{Introduction}
%
% ----------------------------------------------------------

\subsection{Side-Information Source Coding}

%Given a source of two correlated random variables $X$ and $Y$, the minimal achievable rate of joint encoding is $H(X,Y)$ bits. The encoder can first encode $Y$ using $H(Y)$ bits and then encode $X$ using a complement of $H(X|Y)=H(X,Y)-H(Y)$ bits. Surprisingly, this rate is still achievable if the encoders do not operate jointly but separately \cite{slep73}. In particular, the encoder can encode $X$ in $H(X|Y)$ bits such that the decoder can decode it with arbitrarily low probability of error, given that only the decoder has access to (a lossless encoding of) $Y$. In this last setting, $Y$ is called the \emph{side information} and the encoding of $X$ is known as \emph{side-information source coding}.

Given a source of two correlated random variables $X$ and $Y$, the minimal achievable rate of encoding of $X$ is $H(X|Y)$ when $Y$ is given losslessly to the decoder. Surprisingly, this rate is also achievable when $Y$ is known only to the decoder, not to the encoder \cite{slep73}. In this setting, $Y$ is called the \emph{side information} and the encoding of $X$ is known as \emph{side-information source coding}.

The construction of efficient side-information source coding schemes is a difficult problem \cite{zhao03}. Recently, turbo codes have shown to be good candidates for this coding application \cite{aaro02}.

\subsection{The Turbo Principle}

Turbo coding was first introduced in 1993 by Berrou et al. \cite{berr93}. Since then it has been intensively studied and has proved to approach the Shannon limit closer than any other known forward error correcting code. The efficiency of the turbo codes is due to the use of an iterative process at the decoder side and the presence of an interleaver at the encoder side, which adds randomness-like effect to the code.

Our motivation for studying side-information coding in general, and turbo codes specifically, is described next.

\subsection{Quantum Key Distribution}

Quantum key distribution (QKD), also called quantum cryptography, allows two parties, Alice and Bob, to share a secret key that can be used for encrypting messages using a classical cipher, e.g., the one-time pad. The main interest of such a key distribution scheme is that any eavesdropping is, in principle, detectable as the laws of quantum mechanics imply that measuring a quantum state generally disturbs it.

To share a secret key, a few steps must be performed. First, quantum states are sent from Alice to Bob, or vice-versa, on the so-called quantum channel. This process gives the two parties correlated random variables, $X_A$ and $X_B$. Then, using a classical public authenticated channel, Alice and Bob compare a sample of the transmitted data, from which they can determine an upper bound on the amount of information a possible eavesdropper may have acquired. Finally, they distill a common secret key $K$, which is conventionally a function of $X_A$.

Secret key distillation \cite{maur93} usually involves two steps. In the first step, called \emph{reconciliation}, Alice and Bob exchange information over the public authenticated channel in such a way that Bob can recover $X_A$ knowing $X_B$. The exchanged information is considered known to an eavesdropper. The second step consists in applying a \emph{privacy amplification} protocol \cite{benn95:pa} to wipe out  the enemy's information on both quantum and classical transmissions, at the cost of a reduction in the key length. This reduction is roughly equal to the number of bits known to an eavesdropper \cite{benn95:pa}.

It thus appears clearly that reconciliation should not give more information than necessary on $X_A$, otherwise resulting in a penalty in the key length.
Hence, the interest of investigating the use of (efficient) turbo coding in this context.
% It is therefore important to be able to encode $X_A$ using the least number of bits, with $X_B$ side-information known only at the receiver.

\subsection{Problems with Interactive Reconciliation}

An additional motivation for using turbo codes in the scope of QKD lies in that reconciliation is traditionally performed using interactive protocols, such as Cascade \cite{bras93}. While they are perfectly suited to discrete QKD protocols, such as BB84 \cite{benn84}, they suffer from both practical and fundamental problems when used for continuous-variable QKD protocols \cite{gros03}. For a given number of reconciliation bits transmitted from Alice to Bob, interactive protocols impose an additional penalty on the key length over one-way protocols, due to the information leaked from the reconciliation bits originating from Bob. Furthermore, the evaluation of this leaked information depends on the particular eavesdropping strategy, which rules out the use of this method when no assumption on the enemy's side may be made. More details are given in Sec.~\ref{recContQKD}.

Replacing interactive reconciliation protocols by efficient side-information coding is thus another strong motivation for studying this application of turbo coding.

% ----------------------------------------------------------
%
\section{Turbo Coding with Side Information}
%
% ----------------------------------------------------------

\subsection{Turbo Encoder and Turbo Decoder}

A turbo encoder is a parallel concatenation of two, or more, constituent codes separated by one, or more, interleavers. The constituent codes are usually two identical recursive systematic convolutional codes. The input sequence to be encoded is divided into blocks of length $N$. Each block is encoded by the first encoder and interleaved before passing through the second encoder. In channel coding, the systematic output of the first encoder, along with the parity check bits of both encoders are transmitted through the channel. Such a scheme usually uses rate half constituent encoders, so the overall rate is one third. The rate can be increased by puncturing a fraction of the parity bits.

The turbo decoder consists of two, or more, Soft-In Soft-Out (SISO) maximum likelihood decoders. Those decoders operate in parallel, passing extrinsic information to one another in an iterative way. The error rate is lowered after each iteration but the gain in bit error rate decreases as the number of iterations increases, so for complexity reasons the decoder typically performs between 6 and 20 iterations.

Two families of decoding algorithms are commonly used in turbo decoding: Soft Output Viterbi Algorithms (SOVA) and Maximum A Posteriori (MAP) algorithms. The MAP algorithm \cite{bahl74} is more efficient but more complex than the SOVA. However, simplified versions of this algorithm such as MAX-Log-MAP and Log-MAP perform almost as well with a reduced complexity.

%Each encoder can be represented by a trellis diagram in which each node represents a state of the shift registers and each branch represents a transition between two states, depending on the value of the input bit. The decoding algorithm must find the path through the trellis corresponding to the correct input sequence.Descriptions of the different algorithms can be found in \cite{bahl74, wood00, barb96}.

\subsection{Application to Side-Information Source Coding}
In the turbo coding principle, the systematic output of the first component encoder is sent through the channel together with the parity bits from the two encoders. Turbo coding can be used for side-information source coding if we consider the input bit sequence as the random variable $X$, the systematic output of the channel as the side information $Y$, and the parity bits from the two encoders as the information provided to recover $X$ from $Y$.

Thus, in practice, $Y$ is a noisy version of $X$ that is known by the receiver, $X$ is encoded with a turbo encoder by the emitter but only the parity bits are transmitted, and the receiver uses those parity bits and $Y$ to recover $X$ by turbo decoding. To achieve a transmission rate close to the Slepian-Wolf limit, an appropriate puncturing pattern must be used to transmit only a fraction of the produced parity bits.

% ----------------------------------------------------------
%
\section{Reconciliation for Continuous-Variable QKD}
%
% ----------------------------------------------------------
\label{recContQKD}

Gaussian-modulated QKD protocols using coherent states have shown to deliver higher secret bit rates than those based on single photons while using
standard telecom optical components \cite{gros03}. Since they produce continuous variables (i.e., $X_A$ and $X_B$ are correlated Gaussians), a reconciliation procedure adapted to this situation must be used. We here assume that the variable $X_A$ is converted into bits, as described in \cite{vana01} and implemented in \cite{gros03}.

Without going into the details, each instance of $X_A$ is transformed into $m$ bits, making $m$ $l$-bits strings $\{S_i\}_{i\in \{1\dots m\}}$, each called a \emph{slice}, when a run of the QKD protocol produces $l$ instances of Gaussian variables. These $ml$ bits will serve as input to the privacy amplification protocol. On his side, Bob needs to determine Alice's bit values. For this, he calculates his best estimate of $S_i$ given the $l$ values of $X_B$, thus producing the $l$-bit string $\tilde{S}_i$ for each $i$. (\footnote{Actually, the slices are corrected sequentially, for $i=1\dots m$, so that the estimation of $\tilde{S}_i$ can also depend on the knowledge acquired from the previous corrected slices $S_j$, $j<i$ \cite{vana01}.}) Using a binary reconciliation protocol, Bob then recovers $S_i$ given his knowledge of $\tilde{S}_i$.

Even though we started from continuous variables, we thus reach a situation where Alice and Bob need to reconciliate the binary string $S_i$, given that Bob knows the correlated binary string $\tilde{S}_i$. The two strings are related by the error rate $e_i$, that is the probability that a bit of $S_i$ is not equal to the corresponding bit in $\tilde{S}_i$. Overall, the reconciliation produces $H(S_{1\dots m})$ uniform bits by disclosing $\sum_{i=1\dots m} f(e_i)$ bits, with $f(e)$ the number of bits needed to encode a $l$-bit string given that the decoder knows a correlated string with bit error rate $e$. The net result is thus $H(S_{1\dots m})-\sum_{i=1\dots m} f(e_i)$.

\subsection{Binary Reconciliation}

Let us discuss the different options for the binary reconciliation of a given slice with error rate $e$. Of course, it is always possible to encode $S$ using $l$ bits, so that $f(e)\leq l$.

Using a interactive reconciliation protocol such as Cascade \cite{bras93} implies that Alice and Bob exchange parities of various subsets of their strings. After running Cascade, Alice and Bob have disclosed $RS$ and $R\tilde{S}$ for some binary matrix $R$ of size $d \times l$. They thus have communicated the parities calculated over $d$ identical subsets of bit positions. The matrix $R$ and the number $d$ of disclosed parities are not known beforehand but are the result of the interactive protocol, depending on the diverging parities encountered. For Cascade, $d \approx l(1+\xi)h(e)$, where $h(e)=-e \log e - (1-e) \log (1-e)$ and $\xi$ is some small overhead factor $\xi \ll 1$.

In the case of balanced bit strings (i.e., the probabilities of 0 and 1 are the same), the parities $RS$ give Eve $d$ bits of information on $S$, but $R\tilde{S}$ does not give any extra information since it is merely a noisy version of $RS$, or stated otherwise, $S \to RS \to R\tilde{S}$ is a Markov chain.

However, in the more general case where we need to take into account that Eve gathered in $E$ some information on both $S$ and $\tilde{S}$ by eavesdropping on the quantum channel, $S|E \to RS|E \to R\tilde{S}|E$ does not necessarily form a Markov chain. Instead, the actual number of bits disclosed during reconciliation, namely $I(RS,R\tilde{S};S|E)$, must be explicitly evaluated. This quantity is in general larger than $d$, therefore adding an extra cost due to interactivity.

Furthermore, it is unfortunately impossible to evaluate this quantity without making an assumption on the eavesdropping strategy, since we need to explicitly express the variable $E$. In \cite{gros03}, this quantity was calculated for the most general assumption within the scope of that paper (i.e., assuming any individual Gaussian eavesdropping strategy). However, beyond this assumption, the calculation loses its validity.

To remove any assumption, a possibility is to upper bound the number of disclosed bits as if both parties disclosed independent information, that is, $I(RS,R\tilde{S};S|E) \leq 2d \approx 2l(1+\xi)h(e)$. This is unfortunately too expensive in practice, except when $e$ is small, and causes the secret key rate to vanish if this worst-case measure is taken for all slices.

Another option for reconciliation is of course to use side-information source coding, as we will do in Sec.~\ref{secApp}. This provides a non-interactive reconciliation protocol that has the advantages of being independent of the eavesdropping strategy and free of interactivity cost.

% ----------------------------------------------------------
%
\section{Application to an Efficient QKD Protocol}
\label{secApp}
%
% ----------------------------------------------------------

%Alice enters her string in a turbo encoder and sends the generated parity bits to Bob. Those bits are supposed to be transmitted error-free and Bob uses them with a turbo decoder and his string as side-information, to recover the string of Alice.

\subsection{Settings}
We used a turbo code as a binary reconciliation protocol in the continuous-variable QKD protocol described in \cite{gros03}.

The component encoders are two 16-state duo-binary recursive systematic convolutional encoders with generator polynomials (23, 35) \cite{berr03}. The interleaver is a variation of the odd/even interleaver presented by Barbulescu \cite{barb96}. Our interleaver separates the information bits into two groups: group 1 contains bits whose corresponding parity bits have been sent and group 2 contains bits whose corresponding parity bits have been punctured. The two groups are interleaved separately in a pseudo-random manner. Then, the bits are rearranged so that when the second component encoder computes his parity bits, those corresponding to bits from group 1 are punctured in priority, and vice versa. This procedure prevents us from transmitting the two parity bits from one bit, while another bit has none of his parity bits transmitted.
The puncturing pattern depends on the estimated error rate and is chosen to minimize the number of bits sent to Bob.

The decoding algorithm is the Log-MAP algorithm, which is similar to the MAP algorithm but operates in the log-domain. We applied a scheme proposed by Fujii et al. \cite{fuji01}, which consists of weighting the extrinsic information exchanged by the two decoders by a factor depending on whether or not the corresponding parity bit has been received for this bit. We performed 18 iterations with block size $N=10000$.

\subsection{Results}
Each binary string $S_i$ was reconciliated using one of the $3$ following strategies, depending on the estimated error rate $e_i$:
\begin{itemize}
\item if $e_i>15\%$, the string was completely revealed, disclosing $l$ bits of information,
\item if $e_i<0.8\%$, an interactive error correction protocol (Cascade) was preferred and the number of disclosed bits was counted independently for Alice and Bob,
\item otherwise, the turbo coding scheme described above was used.
\end{itemize}

In Table~\ref{mytable}, our results are compared with those of \cite{gros03} based on the use of reverse reconciliation with estimate of the interactivity cost under assumptions. An example of the processing of each slice is given in Table~\ref{mytable2}. For higher losses, the gain on the interactivity cost more than compensates for the higher number of parity bits revealed by a turbo code.

\begin{table}[ht]
\begin{center}
\begin{tabular}{|c|c|c|c|}
\hline
Modulation & Losses & Results from \cite{gros03} & Cascade and Turbo \\
Variance & $(\mathrm{dB})$ & rate $(\mathrm{kbs}^{-1})$ & Code rate $(\mathrm{kbs}^{-1})$ \\
\hline
41.7 & 0 & 1690 & 1605 \\
38.6 & 1.0 & 470 & 450 \\
32.3 & 1.7 & 185 & 209 \\
27 & 3.1 & 75 & 81 \\
\hline
\end{tabular}
\caption{\emph{Net secret key rate with modulation frequency of 800 $kHz$.}}
\label{mytable}
\end{center}
\end{table}
\begin{table}[ht]
\begin{center}
\begin{tabular}{|c|c|c|c|c|}
\hline
Slice & Estimated & Binary Correction & Bits & Shannon Limit \\
Number & BER $e_i$ (\%) & Protocol & Disclosed & $h(e_i)$ \\
\hline
1 & $49.68$ & Full disclosure & $l$ & $0.99$ \\
2 & $34.89$ & Full disclosure & $l$ & $0.93$ \\
3 & $6.38$ & Turbo code & $0.46 l$ & $0.34$ \\
4 & $0.02$ & Cascade & $2 \times 0.005 l$ & $0.0027$ \\
5 & $6 \times 10^{-12}$ & Cascade & $2 \times 0.004 l$ & $3 \times 10^{-10}$ \\
\hline
\end{tabular}
\caption{\emph{Disclosed bits for each slice, corresponding to the 2nd row of Table~\ref{mytable}.}}
\label{mytable2}
\end{center}
\end{table}

% ----------------------------------------------------------
%
\section{Conclusion}
%
% ----------------------------------------------------------
We have shown that decoupling reconciliation and eavesdropping analysis in continuous-variable QKD protocols by using turbo codes allows close, if not better, results than by using Cascade and an evaluation of interactivity costs under assumptions.
%We have shown that the use of turbo codes in continuous-variable QKD protocol leads to close, if not better, results than using an interactive reconciliation protocol such as Cascade, while completely decoupling reconciliation and eavesdropping analysis. 
Furthermore, this opens the way to enhancing the secret key rate for lossy (long-distance) transmissions, for which the interactivity
seems to play a critical role.

%\section*{Acknowledgment}
%\addcontentsline{toc}{section}{Acknowledgment}
%We thank Jean Cardinal for suggesting this research topic and for many fruitful discussions.

\bibliography{qit,cit}

% ----------------------------------------------------------
% ----------------------------------------------------------

\end{document}